\documentclass[a4paper,11pt]{article}

\usepackage[utf8]{inputenc}
\usepackage[T1,T2A]{fontenc}
\usepackage{amsmath,amsfonts,amssymb}

\usepackage{cite}

\usepackage{graphicx}
\usepackage{wrapfig}
\usepackage{subfig}
\usepackage{hyperref}

\numberwithin{equation}{section}

\newcommand{\lb}[0]{\left(}
\newcommand{\rb}[0]{\right)}

\newcommand{\Mpl}[0]{M_\text{Pl}}
\newcommand{\Gev}[0]{\text{GeV}}

\begin{document}

\begin{titlepage}

\begin{center} {\LARGE \bf Dark matter production and reheating  via direct inflaton couplings:  collective effects} \end{center}

\vspace{1cm}

\begin{center}
  {\bf Oleg Lebedev\(^*\), Fedor Smirnov\(^{\dagger,\ddagger}\),
  Timofey Solomko\(^\dagger\), Jong-Hyun Yoon\(^*\)}
\end{center}
  
\begin{center}
  \vspace*{0.15cm}
  \it{\({}^*\)Department of Physics and Helsinki Institute of Physics,\\
  Gustaf H\"allstr\"omin katu 2a, FI-00014 Helsinki, Finland}\\
  \vspace*{0.15cm}
  \it{\({}^\dagger\)Saint Petersburg State University, 7/9 Universitetskaya nab.,\\
  Saint-Petersburg, 199034, Russia}\\
  \vspace*{0.15cm}
  \it{\({}^\ddagger\)ITMO University, Kronverksky pr. 49,\\
  Saint-Petersburg, 197101, Russia}
\end{center}
  
\vspace{2.5cm}

\begin{center} {\bf Abstract} \end{center}

\noindent We study scalar  dark matter production and reheating via renormalizable inflaton couplings,
which include both quartic and trilinear interactions. These processes often depend crucially on collective effects
such as resonances, backreaction and rescattering of the produced particles. To take them into account, we perform lattice simulations and map out parameter space producing the correct 
(non--thermal) dark matter density. We find that the inflaton--dark matter system can reach a quasi--equilibrium state during preheating 
already at very small couplings, in which case the dark matter abundance becomes independent of the inflaton--dark matter coupling and is described by a universal formula.
Dark matter is readily overproduced and even
tiny values of the direct inflaton
  couplings can be sufficient to get the right composition of the Universe, which reaffirms their importance in cosmology.

\end{titlepage}

\tableofcontents

\section{Introduction}

 The stage of particle production  after inflation represents  
 one of the essential  phases in   the Early Universe evolution \cite{Linde:1990flp}. In this epoch, both the 
 Standard Model (SM) fields and dark matter (DM) are generated.
 The underlying mechanisms could be simple perturbative decay of the inflaton
 or more involved non--perturbative particle production by an oscillating background.
 The abundance of dark matter is sensitive to the production mechanism and the corresponding couplings 
 unless it thermalizes thereby erasing its ``memory''.

 In this work, we study scalar dark matter production and reheating within the simplest  
 framework that  contains only renormalizable couplings.
 The unique  renormalizable gauge--invariant interaction between the Standard Model and the inflaton $\phi$ has the form
   \begin{equation}
 V_{\phi h} = {1\over 2} \lambda_{\phi h} \phi^2 H^\dagger H +  \sigma_{\phi h} \phi H^\dagger H \;.
 \end{equation}
 It is thus natural to expect that these couplings play a leading role in producing the SM particles after inflation, i.e. reheating.
 If dark matter is a scalar field $s$, analogous renormalizable terms 
   can be written down for the interaction between $\phi$ and $s$. Although a similar statement applies to the Higgs--DM interaction
   (see \cite{Lebedev:2021xey} for a review), in the present work we focus on non--thermal dark matter and  assume that such couplings
   are negligible. The above interactions are sufficient to fully describe both dark matter production and reheating.
 
 After inflation ends, the inflaton oscillation epoch sets in.
 During this epoch, particle production can be very 
 efficient. Indeed, couplings of the above type can lead to parametric \cite{Kofman:1994rk,Kofman:1997yn,Greene:1997fu} or tachyonic \cite{Felder:2000hj,Dufaux:2006ee}  resonance
 signifying explosive particle production. Certain aspects of this process can be described analytically with the help of semi--classical methods.
 However, it is notoriously difficult to properly account for backreaction and rescattering of the produced particles. It is thus often necessary to resort
 to classical lattice simulations \cite{Khlebnikov:1996mc,Prokopec:1996rr}. To this end, we use the numerical tool  LATTICEEASY~\cite{Felder:2000hq}.  We find that collective effects make a crucial impact on the 
 dynamics of the system and the resulting dark matter abundance. In particular, the produced DM quanta can efficiently scatter against the inflaton background thereby 
 destroying it and possibly bringing the system to quasi--equilibrium.  This regime is highly non--linear and can only be handled adequately with the help of lattice simulations.
 Nevertheless, the resulting DM abundance can be expressed in a transparent and  intuitive form.

We explore   various combinations of the quartic and trilinear couplings making the dominant contribution to the dark and visible matter production, respectively.
Generally, these are produced via different mechanisms and at different times. 
We focus on $\phi^2$ and $\phi^4$ inflaton potentials during the preheating stage, which are not directly related to the inflationary potential at large field values. 
Our aim is to  map out parameter space leading to the correct DM relic abundance, 
under the  assumption that dark matter is much lighter than the inflaton.
 
 The question we try to address in this work is paramount: renormalizable gauge--invariant couplings are generally present and cannot be forbidden unless additional symmetries are invoked. 
 Even their tiny values can make a significant  impact, especially 
 when boosted by the collective effects. While some aspects of the model have been studied before (see, e.g.\,\cite{Heikinheimo:2016yds,Heurtier:2017nwl}), a proper account of collective effects has been lacking.

 To simplify the formulas, in what follows, we employ the Planck units, 
\begin{equation}
M_{\rm Pl} =1 \;,
\end{equation}
where \(\Mpl=(8\pi G)^{-1/2}\) is the reduced Planck mass.

\section{The minimal set--up}

In this work, we consider the possibility that dark matter is {\it non--thermal}. In this case, the production mechanisms
for both dark and SM matter play a crucial role in getting the right composition of the Universe. 
Unlike for WIMPs, the  DM abundance retains the memory of the production mode and thus is sensitive to
the DM--inflaton and Higgs--inflaton couplings. The goal of this work is to delineate parameter space 
leading to the correct relic abundance, taking into account the important collective effects.

Suppose that inflation is driven by a real scalar $\phi$ with mass $m_\phi$.
The only renormalizable inflaton  couplings to the Standard Model are 
 \begin{equation}
 V_{\phi h} = {1\over 4} \lambda_{\phi h} \phi^2 h^2 + {1\over 2} \sigma_{\phi h} \phi h^2 \;,
 \end{equation}
 where we have assumed the unitary gauge for the Higgs field $H$,
  \begin{equation}
H(x)= {1\over \sqrt{2}} \left(
\begin{matrix}
0\\
h(x)
\end{matrix}
\right)\;.
\end{equation}
 Such interactions are expected to play the main role in reheating the Universe. The simplest option 
 to account for dark matter is to introduce a real scalar $s$ with bare mass $m_s$ and impose the parity symmetry $s\rightarrow -s$, which makes it stable.  The renormalizable DM couplings to the inflaton 
 are then given by  
  \begin{equation}
 V_{\phi s} = {1\over 4} \lambda_{\phi s} \phi^2 s^2 + {1\over 2} \sigma_{\phi s} \phi s^2 \;.
 \end{equation}
 In this work, we are interested in DM produced directly by the inflaton, so we neglect a possible $\lambda_{sh}h^2 s^2/4 $ coupling. The latter would  contribute to DM production via the freeze--in
 mechanism  \cite{McDonald:2001vt,Hall:2009bx}  in the Higgs thermal bath.  The resulting contribution to the DM abundance is negligible if \cite{Lebedev:2019ton}
   \begin{eqnarray}
   && \lambda_{s h} < 2 \times 10^{-11} ~~{\rm for }~~m_s \gtrsim m_h \;, \nonumber \\
   && \lambda_{s h} <  10^{-11}\; \sqrt{ {\rm GeV} /m_s}  ~~{\rm for }~~m_s \ll m_h \;.
   \end{eqnarray}
This coupling gets still generated at 1--loop,
 but is suppressed by a  product of the (small) inflaton couplings as well as the  loop factor, so it can be omitted. Similarly, we neglect dark matter self--interaction.  
 
 In our work, we assume that the scalar potential is stable at large field values.
 Although the current data favor somewhat metastability of the electroweak vacuum \cite{Buttazzo:2013uya}, the uncertainty in the top quark mass  makes it inconclusive. Even if the vacuum is currently metastable,
 the Higgs potential can be stabilized by the Higgs--inflaton coupling above  \cite{Lebedev:2012sy} or via Higgs--inflaton mixing driven by $\sigma_{\phi h}$ \cite{Ema:2017ckf}. A discussion of these issues can be found in \cite{Lebedev:2021xey},\cite{Kost:2021rbi}.

 Dark matter and the Higgs quanta are produced after inflation via their couplings to the inflaton. The production mechanism can be perturbative or
 non--perturbative, depending on the strength of the couplings. In the latter case, 
the collective effects are important: the time--dependent inflaton background can lead to resonant particle 
 production followed by backreaction of these particles on the inflaton. Although some aspects of this phenomenon can be treated analytically,
 we use classical lattice simulations to account for such effects properly.

After inflation, $\phi$ starts oscillating around the  minimum of the scalar potential. 
Locally, the corresponding potential is normally quadratic. However,  in models where
inflation is driven by a non--minimal scalar coupling to curvature, it can also be quartic \cite{Bezrukov:2007ep,Garcia-Bellido:2008ycs}.
This is because the oscillation amplitude far exceeds the inflaton mass
such that the inflaton  can be considered effectively massless.
In what follows, we study both of these possibilities in detail.
Our goal is to analyze    DM production by an oscillating inflaton 
assuming   $m_\phi \gg m_s , m_h$, where $m_h =125$ GeV is the observed
Higgs mass. This (rather common) choice helps highlight the collective effects and avoid unnecessary kinematic factors.

 To avoid stable inflaton relics, the  interaction Lagrangian  should contain a linear in $\phi$ term. 
At late times, the Universe is dominated by the SM radiation, hence a natural possibility
would be to consider $\phi h^2$ as the main driver of reheating. DM can then be produced 
via $\phi^2 s^2$ or $\phi s^2$, depending on the model.  We find that the reverse situation
where the SM and dark matter are 
 generated via $\phi^2 h^2 $  and $\phi s^2$, respectively, results in overabundance of DM 
 making it unrealistic. 
 Altogether, we distinguish the following options for the dominant couplings:
 \begin{eqnarray}
  &&  (a) ~~~V_{\phi s}=\frac{\lambda_{\phi s}}{4}\phi^2 s^2 \, ~~,~~                     V_{\phi h}=\frac{\sigma_{\phi h}}{2}\phi h^2 \;,  \label{a} \\
  &&  (b) ~~~V_{\phi s}=  \frac{\sigma_{\phi s}}{2}\phi s^2   ~~ ~~,~~                     V_{\phi h}=\frac{\sigma_{\phi h}}{2}\phi h^2 \;,  \label{b} \\
  &&  (c) ~~~V_{\phi s}=  \frac{\sigma_{\phi s}}{2}\phi s^2    ~~~~,~~                     V_{\phi h}= \frac{\lambda_{\phi h}}{4}\phi^2 h^2 \;. \label{c}
\end{eqnarray}
The resulting DM abundance depends on the inflaton potential, thus we consider the quadratic and quartic options separately.

 In what follows, we do not distinguish positive and negative $\sigma_{\phi i}$  couplings since the results are essentially the same for both cases, so
 we identify $\vert \sigma_{\phi i} \vert  \equiv \sigma_{\phi i}$. On the other hand, we take $\lambda_{\phi i} \geq 0$ for definiteness.

 \section{Dark matter production via $\phi^2 s^2 $}

Consider case (a), i.e.   the  dominant couplings  are    
\begin{equation}
    V_{\phi s}=\frac{\lambda_{\phi s}}{4}\phi^2 s^2 ~~,~~                     V_{\phi h}=\frac{\sigma_{\phi h}}{2}\phi h^2.
\end{equation}
In this case, DM is produced shortly after the end of inflation when the coherent oscillations 
of the inflaton zero--mode are still intense,
while the Higgs and other SM fields result from late time perturbative inflaton decay.

In what follows, we assume that  $\sigma_{\phi h}$ is sufficiently small  such that it  does not affect the resonant DM production.

\subsection{$\phi^2$ preheating potential}

The simplest option is that the inflaton oscillates in a quadratic potential,
\begin{equation}\label{quad_infl}
  V_\phi=\frac{m_\phi^2}{2}\phi^2,
\end{equation}
where the inflaton mass $m_\phi$ is obtained from the curvature of the potential at $\phi=0$. As stressed above,
it is generally unrelated to the potential behaviour in the inflationary regime where the field values are large.
This is particularly important since the quadratic inflationary potential \cite{Linde:1983gd} is disfavored by the current cosmological data \cite{Planck:2018jri}.
\begin{wrapfigure}{r}{0.4\textwidth}
  \begin{center}
    \includegraphics[width=0.38\textwidth]{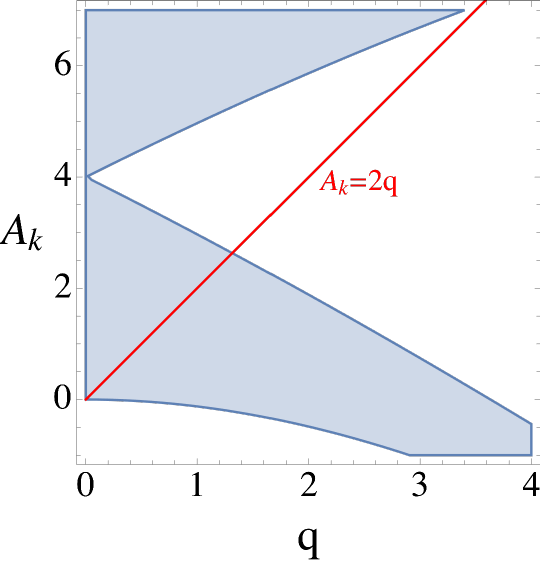}
  \end{center}
  \vspace*{-5mm}
  \caption{\footnotesize Stability chart of the Mathieu equation \eqref{mathieu}.
  In the white regions, the solution grows exponentially, while in the shaded regions it oscillates.
  Below the  \(A_k=2q\) line, the resonance becomes tachyonic.}
  \vspace*{7mm}
  \label{mathieu_chart}
\end{wrapfigure}

 The corresponding equation of motion reads 
\begin{equation}
  \ddot{\phi}+3H\dot{\phi}+m_\phi^2\phi=0\;,
\end{equation}
 where the small Higgs-- and $s$--dependent terms have been neglected.
After a few oscillations, the solution can be approximated by
\begin{equation}\label{infl_sol}
  \phi(t)=\Phi(t)\cos m_\phi t,\quad \Phi(t)=\frac{\Phi_0}{a(t)^{3/2}},
\end{equation}
with  
\begin{equation}
  a(t)=\lb\frac{m_\phi t}{m_\phi t_0}\rb^{2/3} \;,
\end{equation}
where $t_0$ corresponds to the beginning of the preheating epoch, $a=1$. 
The relation between $a$ and $t$ is found by integrating $dt = da/(aH)$ with $H=m_\phi \Phi /\sqrt{6}$, which fixes 
\begin{equation}
m_\phi t_0 = \sqrt{8 \over 3} \; {1\over \Phi_0} \;.
\end{equation} 
We note that Ref.~\cite{Lebedev:2021xey} has studied dark matter production under the simplifying assumption $m_\phi t_0 =1$, which implies $\Phi_0 \sim 1$ in Planck units.
Here we generalize that analysis to arbitrary $\Phi_0$.
 
The equation of motion for the DM field reads
\begin{equation}
  \ddot{s}+3H\dot{s}-\frac{1}{a^2}\nabla^2 s+\frac{\lambda_{\phi s}}{2}\phi^2 s=0 \; .
\end{equation}
In the Hartree approximation, it can be reformulated as a set of decoupled equations for the dark matter 
\(k\)-modes, where  \(k\) is
the comoving  spacial momentum. 
Treating $\phi$ as a background, one finds \cite{Kofman:1997yn}
\begin{equation}\label{mathieu}
  X''_k+\lb A_k+2q\cos 4z\rb X_k=0\;,\quad
\end{equation}
where a small term proportional to  $H^2$ has been neglected and 
the Fourier modes $s_k$ have been traded for $X_k$: 
  $ s_k=a^{-3/2}X_k$. The other quantities are defined by  
\begin{eqnarray}
&&  z\equiv\frac{m_\phi t}{2} ~\;, ~ 
  q \equiv\frac{\lambda_{\phi s}\Phi^2}{2m_\phi^2} ~ \;, ~ 
   A_k \equiv\frac{4k^2}{m_\phi^2a^2} +2q \;.
\end{eqnarray}
Since the time dependence of $A_k$ and $q$ is mild, 
the modes $X_k$ satisfy approximately the  Mathieu equation which describes resonant particle production.

The behaviour of the solution is determined by $q$ and $A_k$: if these belong to the ``stable'' regions,
the solution oscillates, otherwise it grows exponentially. An example is shown in 
Fig.~\ref{mathieu_chart}. In the  shaded regions, the solution is stable. 
Given the initial values of $A_k $ and $q \gg 1$, the system follows a trajectory in this plane, which ends at the origin.
Along the way, $X_k$ goes through regions where it gets amplified until it reaches the last stability zone
at $q \sim 1$. For $k=0$, the trajectory is $A_k=2q$ shown by the red line in Fig.~\ref{mathieu_chart}. 
In the weak coupling regime, the end of {\it parametric} resonance can be identified with the time when\footnote{This depends on the 
convention for $z$. A different convention is used in \cite{Kofman:1997yn}.} 
\begin{equation}
q\simeq 1 \;,
\end{equation}
after which the exponential growth stops.\footnote{Although this statement is $k$--dependent, it works quite well in practice. At $q\ll 1$, the system enters
the narrow resonance regime, however it is inefficient in an expanding Universe and can be neglected for most purposes. }
The amplitude accumulated by that time determines the size of $X_k$ and the corresponding mode occupation number $n_k$ \cite{Kofman:1997yn},
\begin{equation}
n_k = {\omega_k \over 2} \, \left(  {   \vert \dot X_k \vert^2   \over \omega_k^2} + \vert  X_k \vert^2   \right) -{1\over 2} \;,
\end{equation}
with 
 \begin{equation}
\omega_k^2 (t) = {k^2 \over a^2} +  {1\over 2}\lambda_{\phi h} \Phi^2(t) \cos^2 m_\phi t \;.
\label{Xk-EOM}
\end{equation} 
The ``field size'' or the   variance is then computed according to 
\begin{equation}
\langle s^2 \rangle = {1\over (2\pi a)^3 } \int d^3 k \; \vert X_k\vert^2 \simeq  {1\over (2\pi a)^3 } \int  d^3 k \;  {n_k \over \omega_k} \,,
\end{equation}
where $\langle  ... \rangle$ denotes a spacial average.

The above semiclassical treatment of the field $s$ is meaningful if the corresponding occupation numbers are large, $n_k \gg 1$.
It is important to note that $X_k$ is non--zero only if its initial value is non--zero. Such non--trivial boundary conditions are provided by quantum fluctuations
which formally correspond to the initial (renormalized) occupation numbers $n_k =0$ and, hence, non--zero $X_k$.

So far, backreaction of the produced $s$--quanta has been neglected. Yet, it can have an important effect on the evolution of the system.
Indeed, efficient particle production leads to fast growth of $\langle s^2 \rangle $ which induces an effective inflaton mass squared
 ${1\over 2}  \lambda_{\phi s} \langle s^2 \rangle$. When it becomes comparable to $m_\phi^2$, the resonance shuts down.
 At relatively strong coupling, this happens before $q$ reaches 1, so the end of parametric resonance in this case corresponds to \cite{Kofman:1997yn}
 \begin{equation}
  \sqrt{\frac{\lambda_{\phi s}\langle s^2\rangle}{2m_\phi^2}}\sim 1.
\end{equation}

In reality, particle production calculations in our system are complicated by a number of important effects. First, the resonance in an expanding Universe
is not parametric but {\it stochastic}  \cite{Kofman:1997yn}. This is because the Mathieu equation coefficients vary in time which can lead to both 
growth and damping of the amplitude, depending on the phase. Second, the backreaction effects do not simply amount to inducing an effective mass.
They also lead to  rescattering   \cite{Khlebnikov:1996mc} of the DM and inflaton quanta which 
is not captured by the Mathieu equation and 
can fundamentally change the behavior  of the system. 
To take these important effects into account, we employ lattice simulations. A recent related analysis can be found in \cite{Figueroa:2016wxr}.

\subsubsection{Equation of state of the inflaton--DM system}

 \begin{figure}[t]
  \center{\includegraphics[width=0.93\linewidth]{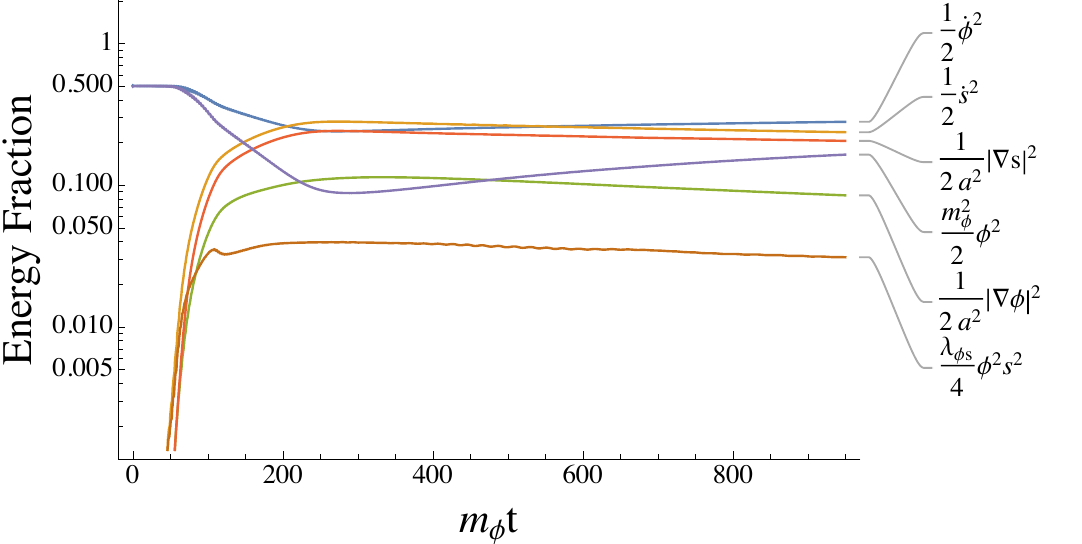}}
  \caption{\footnotesize Contributions of individual terms to the energy balance 
  for $\phi^2$--preheating
  with
  \(\lambda_{\phi s}\approx 2 \times 10^{-6}\), \(\Phi_0\simeq 1\), and
  \(m_\phi\simeq 5 \times 10^{-6}\).  Based on LATTICEEASY simulations.}
  \label{en_ratio_plot}
\end{figure}

The equation of state (EOS) of the system characterized by $w$,
\begin{equation}
p=w \,\rho \;,
\end{equation}
with $p$ and $\rho$ being the pressure and the energy density,
has an important impact on the DM abundance calculation. At weak coupling, the energy density of the Universe is dominated by the non--relativistic inflaton,
while DM behaves as radiation and contributes only a small fraction to the energy balance. In this case, $w\simeq 0$.
At larger couplings, this is no longer true: the DM contribution to $\rho$ is significant and the inflaton itself becomes relativistic. The reason is that the resonance
excites momentum modes far above the inflaton mass, up to
\begin{equation}
k_* \sim m_\phi \, q^{1/4}  \gg m_\phi\;.
\end{equation}
Via rescattering, this momentum gets transferred to the inflaton field making it relativistic. 
Thus, the system goes through a period where $w$ is not far from 1/3 \cite{Podolsky:2005bw}. At a later stage, the momenta redshift to a value around $m_\phi$ such that 
the inflaton becomes non--relativistic and starts dominating again.

To capture these features, we have performed lattice simulations with LATTICEEASY.    
Fig.\;\ref{en_ratio_plot} shows the resulting contributions of different terms in the Hamiltonian to the energy balance at  \(\lambda_{\phi s}\approx 2 \times 10^{-6}\).
We observe that initially the energy is shared equally by the inflaton kinetic and potential terms, while at later times    the DM kinetic and gradient terms become
as important.  This signifies the relativistic behaviour  of the system. In this example, the resonance stops at $m_\phi t \sim 100$ and rescattering makes
the system largely relativistic after $m_\phi t \sim 200$. The typical values of $w$ up to $m_\phi t \sim 1000 $ lie between 0.2 and 0.25. These numbers increase 
somewhat with the coupling. Related lattice studies have recently been performed in \cite{Antusch:2020iyq} with similar results.

\begin{figure}[h!]
\includegraphics[width=0.4999\textwidth]{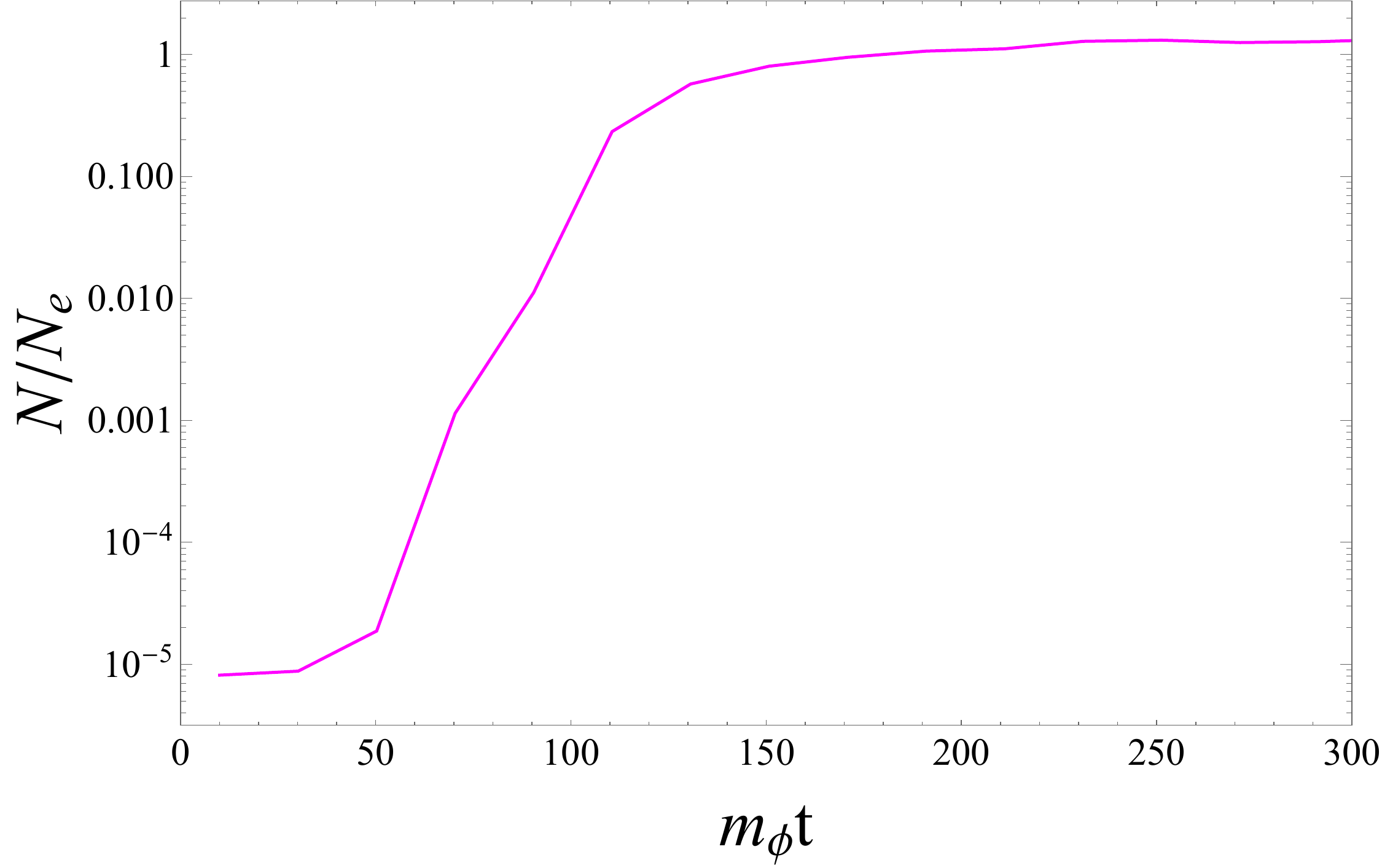}
\includegraphics[width=0.48\textwidth]{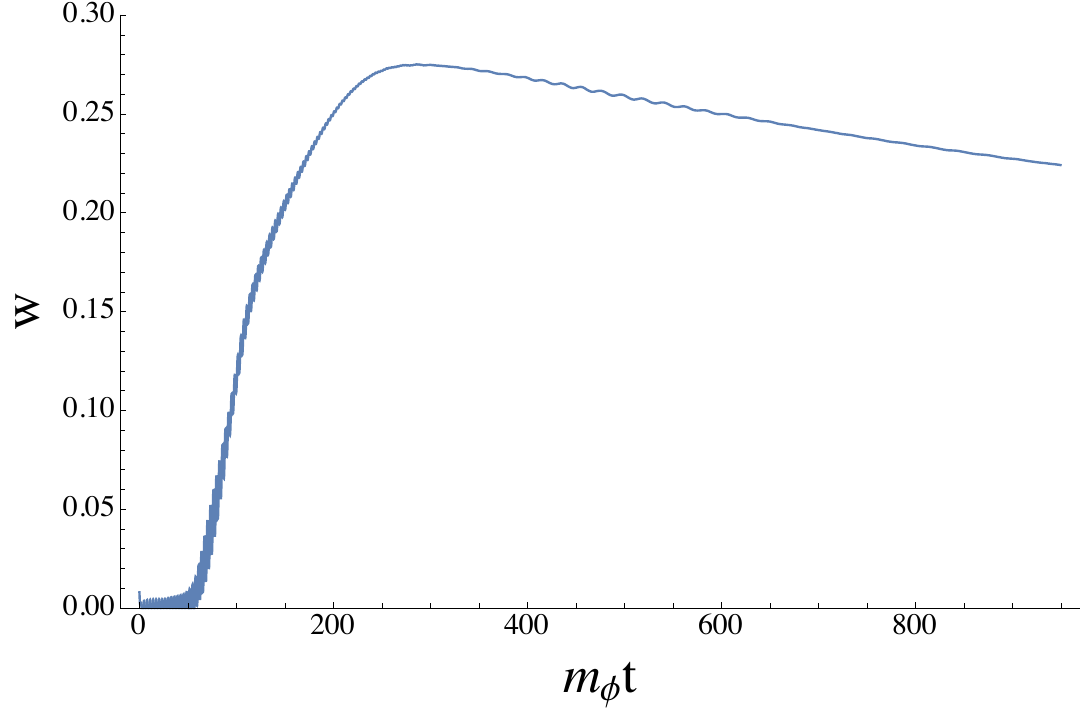}
\caption{{\it Left:} evolution of the DM total particle number $N$ normalized to that at the end of the simulation ($N_e$).  The parameters are  as in Fig.\,\ref{en_ratio_plot}.
{\it Right:} evolution of the equation of state $w$ of the inflaton--DM system.  The figures are produced with LATTICEEASY.} 
  \label{eos_sim_plot}
\end{figure}

The corresponding dark matter particle number and EOS of the  system are plotted in Fig.\;\ref{eos_sim_plot}.  We see that the system remains relativistic for a long period, often beyond the simulation time.
Hence, in our calculations of the DM relic abundance we have to resort to extrapolation. After $w$ reaches its peak, its  evolution with the scale factor $a$ can be approximated by
\begin{equation}
  w_{\rm rel}(a)=\frac{A}{a+B} \;,
  \label{w0}
\end{equation}
 where $A,B$ are constants and the beginning of the simulation corresponds to $a=1$ . For the parameters of Fig.\;\ref{eos_sim_plot}, $A=41$ and $B=106 $, and these coefficients grow with $\lambda_{\phi s}$.
 We expect the above  approximation to be valid in the relativistic regime. When the inflaton becomes non--relativistic, it starts dominating the energy density and $w$ quickly approaches 0.
 Thus, in our calculations we use
\begin{equation}
  w(a)=\frac{A}{a+B} \; \theta (a_* - a),
\label{w}
\end{equation} 
where $a_*$ signifies the onset of the non--relativistic regime, where the typical energy is close to $m_\phi$.

The above extrapolation introduces substantial uncertainty in the DM abundance calculation. To illustrate it, we will present our results for a few choices of $w$.

\subsubsection{Dark matter abundance}

The dark matter abundance is normally expressed in terms of 
    \begin{equation}
    Y = {n \over s_{\rm SM} } ~~,~~ s_{\rm SM} ={2\pi^2 \over 45} \, g_{*s} \, T^3 \;,
    \label{Y}
    \end{equation} 
    where  $n$ is the DM number density, $s_{\rm SM}$ is the SM entropy density at temperature $T$ and 
    $g_{*s}$ is the effective number of SM degrees {} of freedom contributing {} to the entropy. 
    We are interested in non--thermal dark matter at 
     weak $\lambda_{\phi s}$, so the SM entropy is approximately conserved and  $Y$ remains constant
    as long as       the total number of the DM quanta is   constant. 
    The observed value is   \cite{Planck:2015fie}   
     \begin{equation}
    Y_\infty =  4.4 \times 10^{-10} \; \left( {{\rm GeV}\over m_s} \right) \;,
    \end{equation} 
which sets a constraint on the input parameters.

Our task is to compute $Y$ at reheating, after which it stays constant. In order to do so, we need the DM number density at reheating as well as the
reheating temperature $T_R$. The DM number density is a non--perturbative quantity which we determine via lattice simulations, while $T_R$
is computed via late time inflaton decay  $\phi \rightarrow hh$.

The evolution of the system proceeds in  stages. First, DM is produced via parametric resonance, followed by rescattering. Depending on the coupling, 
the system can become relativistic and, at a later stage, return to the non--relativistic state with the energy density dominated by the 
inflaton.\footnote{In the relativistic regime, $s$ still carries a large fraction of the total energy density and  inflaton decay at this stage  
  would generally lead to overabundance of dark matter (see Section 5).} 
When the Hubble rate decreases sufficiently such that it becomes comparable to the   $\phi \rightarrow hh$  decay    rate,
reheating occurs almost instantaneously and 
\begin{equation}\label{reh_cond}
  H_R \simeq \Gamma_{\phi\rightarrow hh} \;,\quad
  \Gamma_{\phi\rightarrow hh}=\frac{\sigma_{\phi h}^2}{8\pi m_\phi},
\end{equation}
where \(H_R\) is the Hubble rate at  reheating and   \(\Gamma_{\phi\rightarrow hh}\)
takes into account 4 Higgs d.o.f. available at high energies.
This, in turn, fixes the reheating temperature via
\begin{equation}
  H_R=\sqrt{\frac{\pi^2g_*}{90}} \, T_R^2 \;,
  \label{HR}
\end{equation}
where $g_* $  is the effective number of SM degrees {} of freedom contributing {} to the energy density.

Requiring the correct DM relic abundance sets a constraint on $\sigma_{\phi h}$, which we may express directly in terms of the simulation output.
 The result depends on the energy balance between the inflaton and DM after preheating. Let us parametrize
\begin{equation}
  \rho_e (s)=\delta \, \rho_e (\phi) \;,
  \end{equation}
  where $\rho_e (s)$ and $\rho_e (\phi)$ are the DM and inflaton energy densities at the end of the simulation.
 Then, the corresponding Hubble rate is given by  $ H_e=\sqrt{1+\delta} \, \sqrt{{\rho_e (\phi)}/{3}}$.
The subsequent evolution of the Hubble rate depends on the equation of state of the system,  
 \(H\sim a^{-3(w+1)/2}\).  Let us for now use the step function approximation,
  \begin{equation}
a_e  \stackrel{\rm { rel}}  \longrightarrow         a_*     \stackrel{\rm nrel} \longrightarrow a_R \;,
\end{equation} 
i.e. between the end of the simulation $(a_e)$ and the onset of the non--relativistic regime $(a_*)$, the system evolves with $w\simeq 1/3$,
while after that till reheating $(a_R)$, it evolves with $w \simeq 0$. 
The parameter $a_*$ can, for instance, be defined by 
\begin{equation}
  \langle E_e (\phi) \rangle \; {a_e \over a_*} \sim m_\phi \;,
\end{equation}
where $ \langle E_e (\phi) \rangle  $ is the average energy of the inflaton quantum   at the end of the simulation, $ \langle E_e (\phi) \rangle = \rho_e (\phi) / n_e (\phi)$.
At reheating, the energy density is dominated by the non--relativistic inflaton, so  $\rho (s)$ can be neglected and 
\begin{equation}
  H_R=\frac{H_e}{\sqrt{1+\delta}} \, \frac{a_e^2}{a_*^2} \, \frac{a_*^{3/2}}{a_R^{3/2}} \; .
\end{equation}
 Solving for $a_R$ and taking $g_* \simeq 107$,  one then finds  
 \begin{equation}
\sigma_{\phi h} \simeq  1.6 \times 10^{-8} \; \sqrt{m_\phi}  \; {H_e^2 \over (1+\delta) \;n_e} \, {a_e\over a_*} ~~\left(   {{\rm GeV}  \over m_s} \right) 
\label{sigma-DM-eq}
\end{equation}
 in Planck units. The simulation outputs $H_e, n_e, a_e, \delta$  and $ \langle E_e \rangle  $,  which thus determine  $\sigma_{\phi h}$. In strong and weak coupling regimes,
 this result can be rewritten directly in terms of the input parameters, as we show below.
\\ \ \\
\noindent {\bf \underline{Strong coupling}.}
The above  formula takes a particularly simple form when the coupling $\lambda_{\phi s }$ is sufficiently strong such that the system reaches quasi--equilibrium by the end
of the simulation. In this case, the energy is almost equally shared by the inflaton and dark matter, i.e. $\delta \simeq 1$.
Then, \(3H_e^2/((1+\delta)n_e)\)
gives the average energy of the DM quantum at the end of simulation. 
Quasi--equilibrium implies that this also equals the average energy of the inflaton quantum, which 
multiplied by \(a_e/a_*\)  yields \(m_\phi\) by definition. 
As a result,  
Eq.\,\eqref{sigma-DM-eq} takes a universal form 
\begin{equation}\label{sigma_simp}
  \sigma_{\phi h}\simeq 5\times 10^{-9}\;m_\phi^{3/2}\lb\frac{\Gev}{m_s} \rb  \;,
\end{equation}
independently of the coupling and initial conditions. This is expected since the ``memory'' is erased in equilibrium. Consequently, the reheating temperature is a function
of $m_\phi / m_s$ only. At strong coupling, the expression for the DM abundance takes a simple form,
\begin{equation}
Y \simeq 0.4\, {\Gamma_{\phi \rightarrow hh}^{1/2} \over m_\phi } \;.
\label{Ystrong}
\end{equation} 

The above results also apply to the $\phi^4$ preheating
potential, as long as $\lambda_{\phi s}$ is sufficiently strong.
 \\ \ \\
 {\bf \underline{Weak coupling}.}
At weaker $\lambda_{\phi s}$, the backreaction and rescattering effects are less significant such that the DM output can be estimated using the theory of  broad 
parametric resonance ($q\gg 1$). In this case, the energy density is dominated by the non--relativistic inflaton,
\begin{equation}
  H=\frac{m_\phi\Phi}{\sqrt{6}},
\end{equation}
while DM gives only a small correction, $\delta \simeq 0$ and $a_e/a_* \simeq 1$. 
DM production stops when the $q$--parameter decreases to the value close to 1, after which the total number of $s$--quanta remains approximately constant.
The DM number
density  produced during the resonance can be estimated using~\cite{Kofman:1997yn}
\begin{equation}
  n\sim \frac{k_*^3}{64\pi^2a^3\sqrt{\pi\mu m_\phi (t-t_0)}}e^{2\mu m_\phi (t-t_0)},
\end{equation}
where \(k_*\equiv\lb\sqrt{\lambda_{\phi s}/2}\, m_\phi\Phi_0\rb^{1/2}\) and \(\mu\) is the
effective Floquet exponent. In practice, $t \gg t_0$ and the $t_0$ dependence can be neglected. 
The end of the resonance corresponds to 
\begin{equation}
 \frac{\lambda_{\phi s}\Phi^2}{2m_\phi^2} \simeq 1 \;,
\end{equation}
which can be solved for $t$ and  the scale factor, $m_\phi t_{\rm end} = \sqrt{ 4 \lambda_{\phi s} \over 3} \; {1\over m_\phi}$ and
$a^3_{\rm end } = {\lambda_{\phi s} \Phi_0^2 \over 2 m_\phi^2} $.
 After that, (almost) no DM gets produced and $H^2/n$ remains constant.
One thus finds
\begin{equation}
  \sigma_{\phi h}\sim 2\times 10^{-6} \;\sqrt{\frac{m_\phi \, \Phi_0}{\lambda_{\phi s}}} \;
    \exp\left(- {4\mu \over \sqrt{3} } {\sqrt{\lambda_{\phi s}} \over m_\phi}  \right)\;
  \left( \frac{\rm GeV}{m_s}\right) \;.
\end{equation}
 This result is exponentially   sensitive to the exact value of the 
 effective Floquet exponent $\mu$, hence only an order of magnitude (at best) of $\sigma_{\phi h}$
 can be  estimated. Here, we assume $\mu \sim 10^{-1}$. 
 \\ \ \\
 \begin{figure}[b!]
  \center{\includegraphics[width=0.75\linewidth]{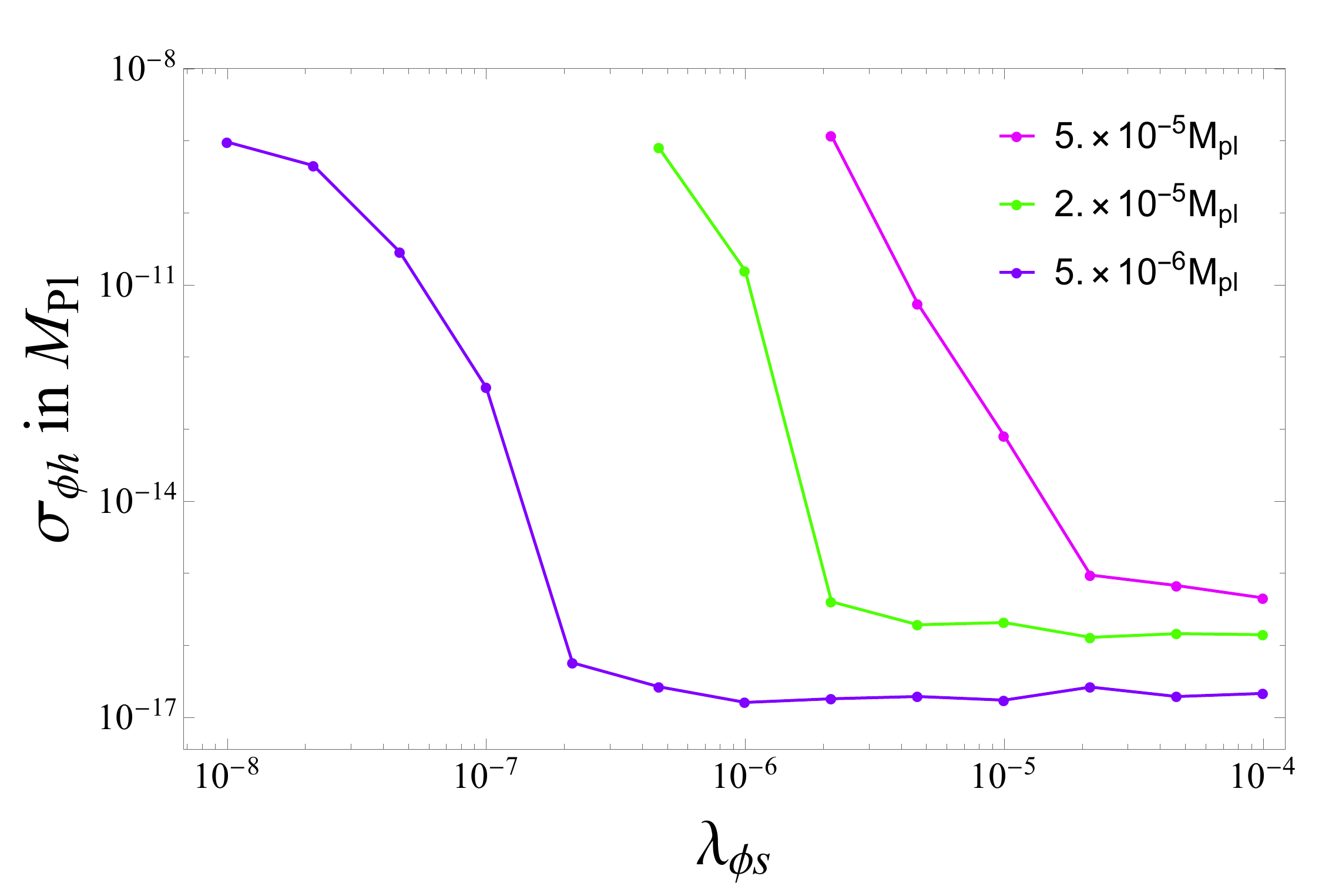}}
  \caption{\footnotesize  \(\sigma_{\phi h}\) vs. \(\lambda_{\phi s}\) 
  producing the correct DM relic abundance in a $\phi^2$ potential  for DM mass $m_s=1\,$GeV and 
  three values of the inflaton mass 
  $m_\phi = 5\times 10^{-5}\, , \, 2 \times 10^{-5}\, , \,
   5\times 10^{-6}$ in Planck units; $\Phi_0 \simeq 1$. 
   The dots on the curves are obtained 
  with
  LATTICEEASY simulations. The area above the curves is ruled out by overabundance of dark matter.}
  \label{sig_lam_phi2_plot}
\end{figure}
 {\bf \underline{Simulations}.}
Our simulation results for three values of the inflaton mass $m_\phi = 5\times 10^{-5}\, , \,
2\times 10^{-5}\, , \,5\times 10^{-6}$  are shown in Fig.\,\ref{sig_lam_phi2_plot}.
   We clearly see the two regimes discussed above: at strong coupling the curves flatten out, while
   at weak coupling $\sigma_{\phi h}$ is exponentially sensitive to $\lambda_{\phi s}$. These results are generally consistent with the analytical considerations presented above,
   although in the weak coupling regime $\sigma_{\phi h}$ is very sensitive to the effective $\mu$ whose value can only be ballparked.
  According to Eq.\,\ref{sigma-DM-eq},
   the values of $\sigma_{\phi h}$ lying above the curves are ruled out by overabundance of dark matter. This is understood intuitively since a larger $\sigma_{\phi h}$ 
   leads to earlier reheating, in which case the DM energy density is not dilute enough.

The range of $\lambda_{\phi s}$ is limited by the following considerations. At large couplings,
loop corrections  affect the inflaton potential, $\Delta V_\phi \sim (\lambda_{\phi s}^2/64\pi^2) \, \phi^4 \, \ln (\phi^2/\phi_*^2)$ with some reference value $\phi_*$. Yet larger $\lambda_{\phi s}$ induce
significant DM self--interaction which can lead to DM thermalization, depending on $m_s$. For example, at $m_s \sim 1\,$GeV, non--thermalization requires $\lambda_s \sim  \lambda_{\phi s}^2/(16 \pi^2)  \lesssim 10^{-4}$, 
while for lighter DM the bound gets stronger \cite{Arcadi:2019oxh}.
At weak couplings corresponding to $q\lesssim 1$, the resonance becomes inefficient and classical simulations
are inadequate. 

\begin{figure}[b!]
  \center{\includegraphics[width=0.75\linewidth]{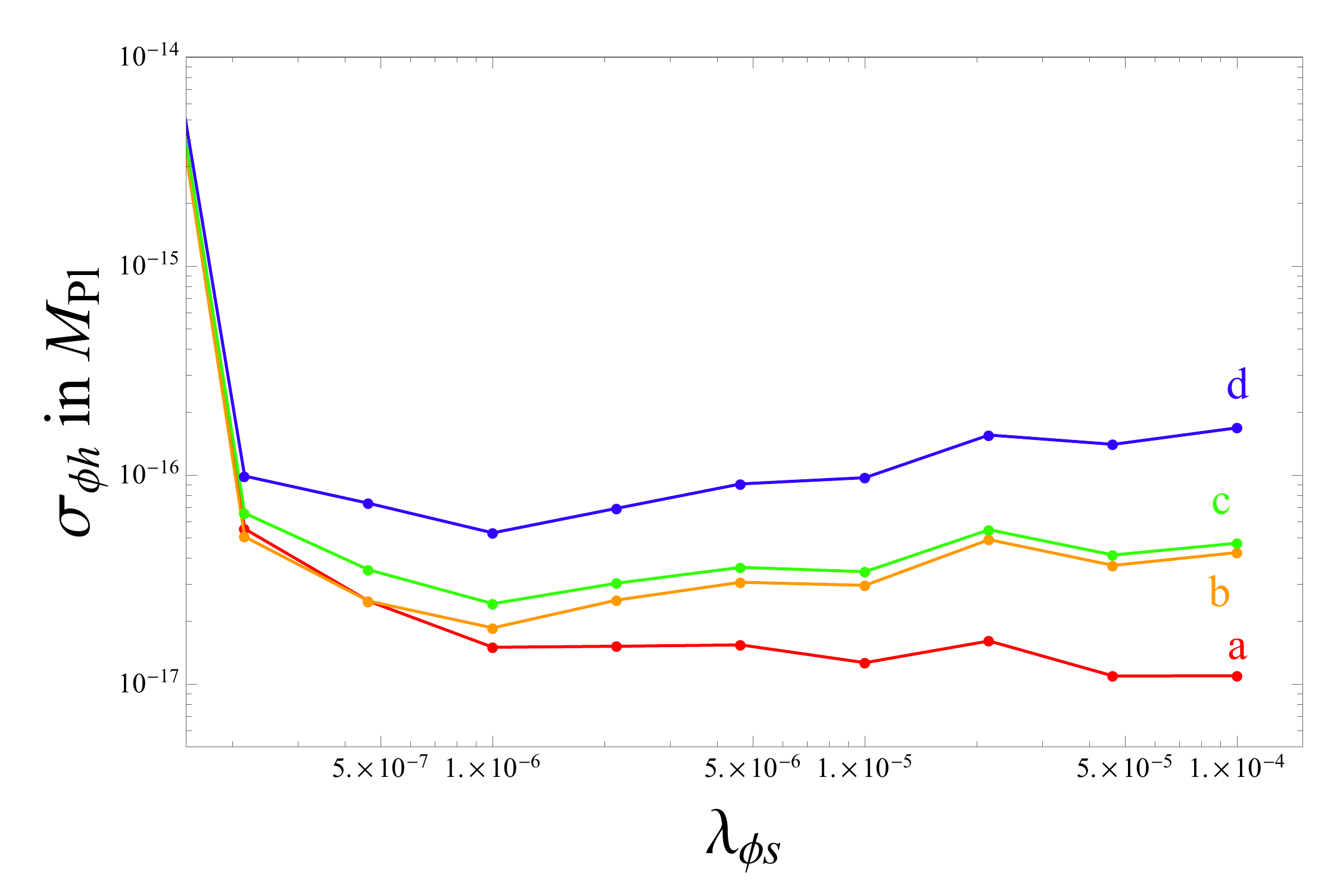}}
  \caption{\footnotesize Sensitivity of $\sigma_{\phi h}$ to the equation of state of the inflaton--dark matter system: (a) $w= A/(a+B)$; (b)
   $w= A/(a+B) \times \theta (a_*-a)$; (c) $w=1/3 \times \theta (a_*-a)$; (d) $w=0$.
 Here $m_\phi = 5\times 10^{-6}$ in Planck units  and the 
  other   parameters are as in Fig.\,\ref{sig_lam_phi2_plot}.
   }
  \label{eos-comp}
\end{figure}

An important source of uncertainty in the calculation of $\sigma_{\phi h}$ at strong coupling is associated with the equation of state
of the inflaton--dark matter system. This is illustrated in Fig.\,\ref{eos-comp}.
In deriving (\ref{sigma-DM-eq}), we have used a simple step--function
\begin{equation}
w(a)= {1\over 3}\, \theta(a_* -a ) \;.
\end{equation}
Although in the radiation--dominated phase $w$ is actually  somewhat below 1/3 (Fig.\,\ref{eos_sim_plot}),
this gives a very similar  result to that with $w(a)$ defined by Eq.\,\ref{w}. If instead we assume that (\ref{w0}) is valid over
the entire relevant range of $a$, the corresponding $\sigma_{\phi h}$ is a few times lower. 
For illustration, we also show another extreme, $w=0$, that gives a somewhat
higher value of $\sigma_{\phi h}$. In the absence of full control over $w$, one should keep in mind the ${\cal O}(1)$
sensitivity of the result to the equation of state of the system.

In the above analysis, we have neglected a few subleading effects. First, dark matter production may continue after the resonance if the inflaton 
zero mode has not been ``destroyed'' by rescattering. We discuss the perturbative DM production below and find that it only gives a small correction
to our result. Second, a significant value of $\sigma_{\phi h}$, such that  $\sigma_{\phi h} \Phi_0 / m_\phi^2 \gg 1$,  would  lead to tachyonic resonance resulting in explosive Higgs production
and may thus affect the inflaton evolution.
We have verified that  this effect is insignificant for the parameter range of interest.

The numerical results presented here can further be  improved by employing  new 
  simulation tools such as Cosmolattice~\cite{Figueroa:2020rrl,Figueroa:2021yhd}.
\\ \ \\
 {\bf \underline{Perturbative DM production}.}
Even in the absence of broad parametric resonance, dark matter can be produced perturbatively by an oscillating inflaton background  \cite{Dolgov:1989us,Traschen:1990sw,Ichikawa:2008ne}.
This process is allowed kinematically if the induced DM mass is below the inflaton mass, $\sqrt{\lambda_{\phi s}/2 } \, \Phi < m_\phi $
or $q<1$, 
and may largely be attributed to  $\phi \phi \rightarrow ss$. The corresponding perturbative reaction rate per unit volume is \cite{Lebedev:2021xey}
  \begin{equation}
\Gamma (\phi \phi \rightarrow ss) \simeq  {\lambda_{\phi s}^2 \over 1024 \pi}  \Phi^4\;,
\end{equation}
neglecting the small narrow resonance effects.
The density of $s$ is then found via the Boltzmann equation,
\begin{equation}
\dot n + 3Hn = 2\, \Gamma (\phi\phi \rightarrow ss) \;.
\end{equation}
Since the energy density is dominated by the non--relativistic inflaton, $\Phi =\Phi_0 \,a^{-3/2}$, the Boltzmann equation 
is easily solved: $a^3 \, n(t) = \alpha - \beta \, a^{-3/2}$, with constant $\alpha,\beta$. 
Imposing the boundary condition $n=0$ at $a=1$, one finds that at $a\gg 1$,
    \begin{equation}
n(t) = {\sqrt{6} \over 768 \pi} \; {\lambda_{\phi s}^2 \Phi_0^3\over m_\phi \, a^3} \;.
\end{equation}
Since the DM production rate drops faster with time than the Hubble rate does, the DM output is dominated by early times and its total number is almost constant. 
Using  $Y = n_R / (7.4 \Gamma_{\phi\rightarrow hh}^{3/2})$ and $a_R^3 = m_\phi^2 \Phi_0^2 / (6 \Gamma_{\phi \rightarrow hh}^2 )$, one finds that the correct DM abundance is produced
for 
  \begin{equation}
\sigma_{\phi h} \simeq 3\times 10^{-6}  \; {m_\phi^{7/2} \over \lambda_{\phi s }^2 \Phi_0 } \; \left(     { {\rm GeV}  \over m_s} \right) 
\end{equation}
in Planck units. For the benchmark values, $m_\phi \simeq 5 \times 10^{-6}$, $\Phi_0 \sim 1$ and $\lambda_{\phi s} \sim 10^{-9}$, the required 
$\sigma_{\phi h}$ is about $10^{-6}$, which is substantially higher than the typical values for resonant production. This is natural since 
$Y \propto  N \,\Gamma_{\phi\rightarrow hh}^{1/2}$ and non--resonant DM production results in a smaller dark matter particle number  $N$.

We can also verify that the perturbative DM production after the resonance gives a small correction to the resonant result. Suppose we are in the weak coupling regime so that
the zero inflaton mode is not destroyed by rescattering. To calculate just the perturbative contribution, we may use the above solution $a^3 \, n(t) = \alpha - \beta \, a^{-3/2}$
with the boundary condition $n=0$ at $q=1$. Then at late times,
\begin{equation}
  n(t) =   {\lambda_{\phi s}^{3/2} \Phi_0^2 \over \sqrt{3} \,128 \pi a^3} \;.
\end{equation}
Again, the production rate decreases faster than $H$, so largest contribution comes from the period right after the resonance.\footnote{We are neglecting a modest Bose--Einstein enhancement of the amplitude due 
to particles produced during the resonance. Their momenta get quickly redshifted below $m_\phi$, in which case the effect is small.}
It is easy to verify that the above $a^3 n(t) $  is significantly smaller than the particle number   produced by the resonance, so the perturbative contribution can be omitted. 
 
For large $\sigma_{\phi h }$, i.e. $\sigma_{\phi h } \Phi_0/ m_\phi^2 \gg 1$, the tachyonic resonance affects the above estimates by making the Higgs production more efficient. 
A detailed analysis of this case is beyond the scope of the present study.

\subsection{$\phi^4$ preheating potential}

Let us now consider dark matter production in the quartic inflaton potential 
\begin{equation}
V_\phi = {1\over 4 } \lambda_\phi \phi^4\;.
\end{equation}
This possibility is realized when the Taylor expansion of the potential is dominated by the fourth derivative,
which can occur, for instance, in models with non--minimal scalar--gravity coupling. 
If one neglects the inflaton and DM masses, the system exhibits approximate conformal invariance so that particle
production has qualitatively different features compared to the $\phi^2$ case~\cite{Greene:1997fu}. 
In a realistic situation, both the inflaton and DM are massive, yet for our purposes they can be treated as massless during the initial 
(non--perturbative) stage of particle production.

The equation of motion for the inflaton reads
\begin{equation}
  \ddot{\phi}+3H\dot{\phi}+\lambda_\phi\phi^3=0.
\end{equation}
It can be rewritten in terms of the conformal time \(\eta\) and rescaled field
\(\psi\),
\begin{equation}
  d\eta=a(t)^{-1} \, dt,\quad
  \psi\equiv a\phi,\quad 
  \end{equation}
and takes the form 
\begin{equation}
  \psi''_\eta +\lambda_\phi\psi^3=0 \;,
\end{equation}
where the prime denotes differentiation with respect to $\eta$.
Here we have omitted 
the \(a''/a\) term, which is justified  after a few inflaton oscillations,
when the averaged  energy-momentum tensor becomes traceless~\cite{Turner:1983he}.
$\psi$ satisfies 
  an elliptic equation whose solution is well known. In terms of $\phi(t)$, the solution takes the form 
\begin{equation}
  \phi (t)=\frac{\Phi_0}{a(t)}\, \text{cn}\lb x,\frac{1}{\sqrt{2}}\rb,\quad
  x\equiv(48\lambda_\phi)^{1/4}\sqrt{t} \, .
\end{equation}
The scale factor satisfies $a(0)=1$ and $a\propto t^{1/2}$ after a few inflaton oscillations.
The solution is an oscillating function in $x$ with a decreasing amplitude. The 
leading   contribution to the Jacobi cosine is given by  \(\cos\frac{2\pi}{T}x\),
 where \(T=\Gamma(1/4)^2/\sqrt{\pi}\).

Similarly to the $\phi^2$ case, we can treat the inflaton as a classical background and write down the equations of motion 
for the dark matter momentum modes $X_k$ (cf.\,Eq.\,\ref{mathieu}) \cite{Greene:1997fu},
\begin{equation}
  X''_k+\lb\kappa^2+\frac{\lambda_{\phi s}}{2\lambda_\phi}\text{cn}^2\lb x,\frac{1}{\sqrt{2}}\rb\rb X_k=0,\quad
  \kappa^2\equiv\frac{k^2}{\lambda_\phi\Phi_0^2} \;,
\end{equation}
where the prime denotes differentiation with respect to $x$.
This belongs to the class of Lam\'e equations. 
 The effective mass term for $X_k$ is periodic in $x$ and thus the Floquet
analysis applies. The behaviour of 
  \(X_k\) is determined by a stability
chart in terms of \(\kappa^2\) and \(q'=\lambda_{\phi s}/(2\lambda_\phi)\). In contrast to  the Mathieu equation case,
 the amplitude growth is controlled by the $ratio$ of
couplings $q^\prime$ which stays constant in time. This is a direct consequence of the (approximate) conformal symmetry of the system.

Since both $\kappa^2$ and $q^\prime$ are constant, the resonance can only end due to backreaction~\cite{Greene:1997fu}.
The produced particles drain energy from
the inflaton field, thereby reducing  its amplitude, and also
induce    an effective inflaton mass, which eventually terminates   resonant particle production.
The resonance in the quartic theory can be
efficient over long time scales even if the resonance is narrow, 
  \(\lambda_{\phi s}/(2\lambda_\phi)\ll 1\). In this case, 
 the Lam\'e equation can be approximated by the Mathieu equation with a $constant$ \(q\ll 1\). 
 The corresponding Floquet exponent is suppressed, yet the exponential amplitude growth is still visible on a long time scale. 
 In the $\phi^2 $
case, however,  the narrow resonance is inefficient due to a decreasing in time  $q$.

Owing to inflaton self--interaction, the inflaton quanta are also produced by the oscillating background. Quantizing the fluctuations
$\delta \phi$, where $\delta \phi = \phi - \langle \phi \rangle$ and $ \langle \phi \rangle$ being the ``classical'' inflaton component,
one finds an analogous Lam\'e equation for the momentum modes of the inflaton fluctuations.
In this case, however, the $q^\prime$ parameter is fixed to be 3. 
This implies, for example, that the inflaton quanta production is more efficient than DM production
if ${\lambda_{\phi s} /( 2 \lambda_\phi}) \lesssim 1/3$.

The 
  backreaction and rescattering effects are crucial for the analysis of the system \cite{Khlebnikov:1996mc,Prokopec:1996rr}, which makes lattice simulations an indispensable  tool. 
   The semi--classical approximation is valid as long as the occupation numbers are sufficiently large, which restricts the range 
   of the couplings that can be studied.
   We find that this is ensured by imposing \(\lambda_{\phi s}\gtrsim 0.5\lambda_\phi\).
   In this case, the system evolution and the DM output can  reliably be computed on the lattice.

\begin{figure}[t!]
  \center{\includegraphics[width=0.75\linewidth]{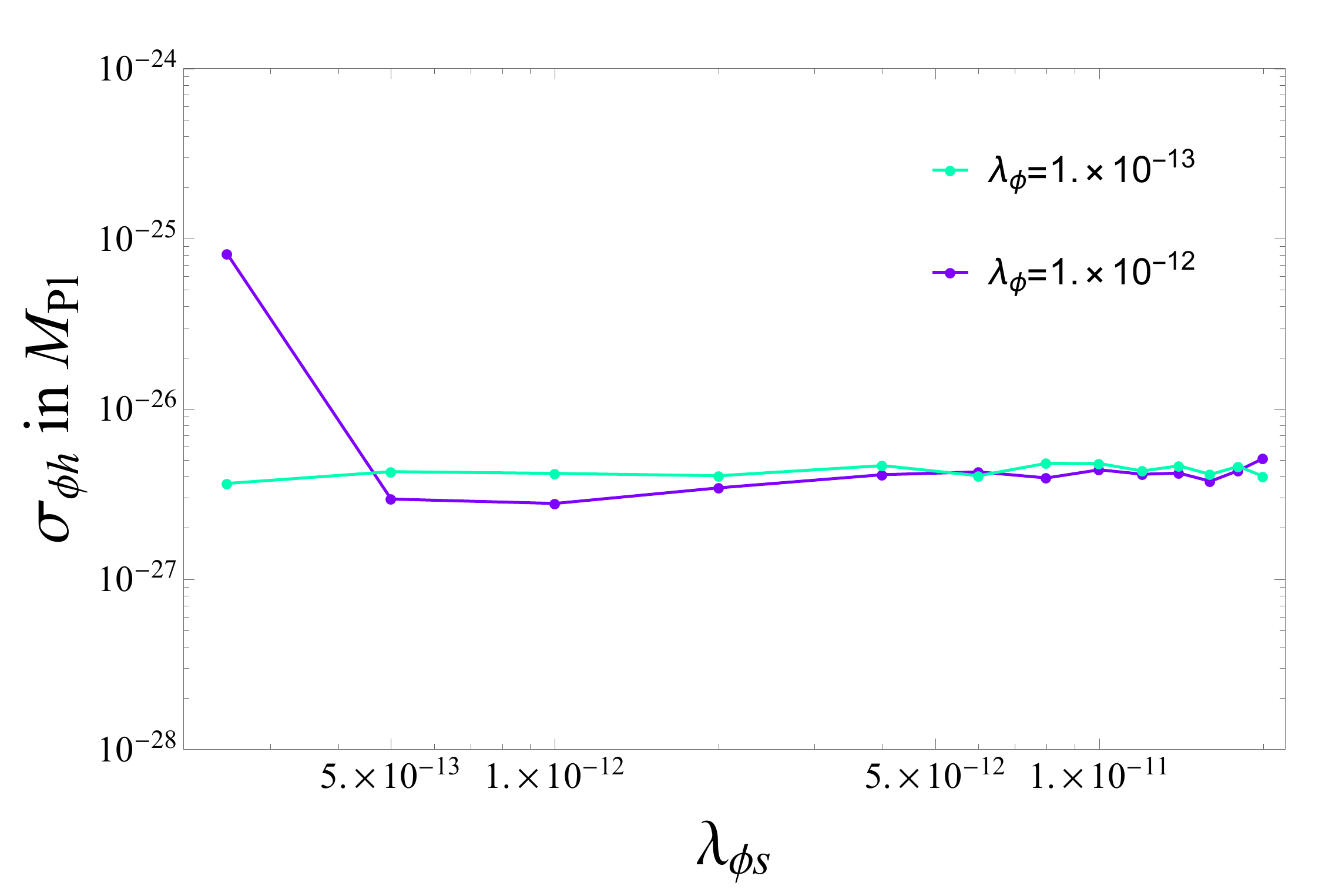}}
  \caption{\footnotesize  \(\sigma_{\phi h}\) vs. \(\lambda_{\phi s}\) 
  producing the correct DM relic abundance in a $\phi^4$ potential  for DM mass $m_s=10\,$keV and 
  $m_\phi = 1\,$TeV; 
    $\Phi_0 \simeq 1.7$. 
   The dots on the curves are obtained 
  with
  LATTICEEASY simulations. The area above the curves is ruled out by overabundance of dark matter.}
  \label{sig_lam_phi4_plot1}
\end{figure}

In the quartic case, dark matter is produced very efficiently. 
  Already at $q^\prime \gtrsim 1 $, about half of the inflaton energy  gets transferred to dark matter and the system reaches the state of 
  quasi--equilibrium by the end of the simulation. 
  It behaves as radiation and the total energy density scales as $a^{-4}$ with the scale factor. 
  The total number of the DM quanta remains approximately constant as soon as rescattering has completed.
  Eventually, the momenta redshift to the point
  that a non--zero $m_\phi$ starts playing a role, after which the inflaton becomes non--relativistic and dominates the energy density. 
  Subsequently, perturbative inflaton decay via $\sigma_{\phi h}$ sets in and the SM radiation gets produced.\footnote{Generally,  the inflaton decay during the relativistic stage leads to overabundance of dark matter, see e.g. Section\,\ref{challenge}. Hence, we assume that the inflaton becomes non--relativistic before it decays. }
  The system goes through the same stages in its evolution as in the $\phi^2$ case, except the relativistic regime lasts much longer.
  The general result (\ref{sigma-DM-eq}) applies and, since the system reaches quasi--equilibrium, the approximate relation (\ref{sigma_simp}),
  namely 
 \begin{eqnarray}
&&  \sigma_{\phi h}\simeq 5\times 10^{-9}\;m_\phi^{3/2}\lb\frac{\Gev}{m_s} \rb  \;,
\end{eqnarray}
holds with good accuracy. This is seen in Fig.\,\ref{sig_lam_phi4_plot1} which clearly shows independence of the result of $\lambda_\phi$ and $\lambda_{\phi s}$. (The outlying point  on the purple curve
does not satisfy \(\lambda_{\phi s}\gtrsim 0.5\lambda_\phi\) and should be discarded.) 

Fig.\,\ref{sig_lam_phi4_plot1} gives a viable example of a system with a light inflaton, $m_\phi = 1\,$TeV, and very light DM, $m_s =10\,$keV. A nontrivial constraint in this case is provided by the 
reheating temperature, $T_R > 4 $ MeV \cite{Hannestad:2004px}.
According to Eqs.\,\ref{reh_cond},\ref{HR} and the above relation, 
\begin{equation}
T_R \propto  m_\phi /m_s \; .
\end{equation} 
As a result, a low $m_\phi$ necessitates very light dark matter. For the above parameters, $T_R \sim 50 $ MeV such that the constraint is satisfied.
We observe that the required $\sigma_{\phi h}$ is much lower than that for the $\phi^2$ preheating potential.
This is natural: in the relativistic regime, the DM and inflaton contributions to the energy density scale the same way. In order to dilute DM, one needs an extended non--relativistic phase, 
which is only possible if $\sigma_{\phi h}$ is tiny: for our parameter choice, $\sigma_{\phi h} \sim 1\,$eV. This makes, in particular, the Higgs production via 
tachyonic resonance  negligible.

The above behaviour persists in a wider range of couplings limited by the following factors.
For a given $\lambda_\phi$, the requirement of having large DM occupation numbers sets a lower bound on $\lambda_{\phi s}$, as mentioned above.
The upper bound on $\lambda_{\phi s}$ is provided by the size of radiative corrections to $\lambda_\phi$ as well as by non--thermalization of the inflaton--DM system.
The first constraint requires $\lambda_{\phi s}^2 / 16\pi^2 \lesssim \lambda_\phi$, while the second depends   on $m_\phi$ and 
can be estimated as $\lambda_{\phi s} < 10^{-7} \, \sqrt{m_\phi / {\rm GeV}}$  \cite{DeRomeri:2020wng}.

\section{Dark matter production via $\phi s^2$}

The traditional and perhaps simplest  option to produce  matter after inflation is to employ the trilinear inflaton couplings \cite{Linde:1990flp},\footnote{The DM field is assumed to have small but non--zero self--interaction  such that the potential is bounded from below. It is unimportant for particle production, yet regularizes the scalar potential at very large field values.}
  \begin{equation}
V_{\phi h} = {1\over 2} \sigma_{\phi h} \, \phi h^2 ~~,~~ V_{\phi s} =   {1\over 2} \sigma_{\phi s} \, \phi s^2 \;,
\end{equation}
which is case (b) in Eq.\,\ref{b}.
The perturbative decay of the inflaton then generates both dark and observable matter. As long as $m_\phi \gg m_h, m_s$,
the Higgs final state should dominate to avoid a dark Universe so that  $ \sigma_{\phi h}  \gg \sigma_{\phi s} $ for non--thermal DM. 
The corresponding decay rates are constant in time, hence they overtake the Hubble rate at some stage which signifies the reheating epoch.

Let us focus on the quadratic inflaton potential $m_\phi^2  \phi^2 /2$ and modest $\sigma$'s such that the resonant effects can be neglected.
In this case, the decay rate of the oscillating inflaton background is equivalent to that of a non--relativistic inflaton quantum,
 \begin{equation}
\Gamma (\phi \rightarrow ss) = {\sigma_{\phi s }^2 \over 32 \pi m_\phi }  ~~,~~ \Gamma_{\rm tot}\simeq  \Gamma (\phi \rightarrow hh)= {\sigma_{\phi h }^2 \over 8 \pi m_\phi } \;,
\end{equation}
where we include all 4 d.o.f. of the Higgs.  The inflaton energy gets converted into the Higgses when $\Gamma_{\rm tot} \simeq H$. Until that time, the effect of the inflaton
decay on its amplitude can be neglected and $\Phi \propto a^{-3/2}$.\footnote{This is clear from the Boltzmann equation $\dot n_\phi + (3H + \Gamma_{\rm tot}) \,n_\phi =0$, whose solution depends on the balance between $3H$ and $\Gamma_{\rm tot}$. It is customary to neglect the factor of 3 when studying reheating, which introduces ${\cal O}(1)$ uncertainty in the final result.}
 The dark matter abundance is determined by the number of DM quanta accumulated up to the reheating 
stage. It is found via the Boltzmann equation
\begin{equation}
 \dot n +3Hn  = 2 \Gamma(\phi \rightarrow ss) \; n_\phi \;,
\end{equation}
where $n_\phi = {1\over 2} m_\phi \Phi^2$ is the inflaton number density.
The boundary condition is $n=0$ at $a=1$. Then, the late time solution reads 
\begin{equation}
n(t) = {2 \sqrt{6} \over 3} \, \Phi_0 \Gamma (\phi \rightarrow ss) \, a^{-3/2} \;.
\end{equation}
Using $Y= n_R / (7.4 \Gamma_{\rm tot}^{3/2})$ and $a_R^{-3/2}= \sqrt{6} \Gamma_{\rm tot}/(m_\phi \Phi_0) $, one gets
\begin{equation}
 \sigma_{\phi h} = 6\times 10^7 \;{\sigma_{\phi s}^2 \over m_\phi^{3/2}}  \; {m_s \over {\rm GeV}} \;.
 \label{sigma-pert}
\end{equation}
As a result, there is a wide range of parameters in which the correct relic density is produced, subject to the constraints $ \sigma_{\phi h}  \gg \sigma_{\phi s};\, \sigma_{\phi h }  \Phi_0 /m_\phi^2 \lesssim 1$.

This perturbative approach breaks down at large $\sigma_{\phi h}$ in which case the Higgs production via tachyonic resonance becomes important.
 However, the boundedness from below of the scalar potential imposes a constraint on the size of $\sigma_{\phi h}$. Indeed, when the potential
 is dominated by $V= m_\phi^2 \phi^2 /2 + \sigma_{\phi h} \phi h^2 /2 + \lambda_h h^4/4$, no run--away direction exists only if
\begin{eqnarray}
\sigma_{\phi h} < \sqrt{2\lambda_h} \, m_{\phi} \; .
\label{sig_cond}
\end{eqnarray} 
Here, as before, we identify  $\sigma_{\phi h}$ with its absolute value. For a realistic Higgs coupling $\lambda_h \sim 10^{-2}$, this constraint is satisfied in all of our previous considerations.
We find that a trilinear coupling subject to the above bound does induce tachyonic resonance. However, it gets quickly shut down by the Higgs self--interaction which generates a significant 
effective Higgs mass. As a result, the inflaton energy transfer to the Higgs is impeded such that the resonance does not have a significant effect on the dynamics of the system.
 We thus find that Eq.\,\ref{sigma-pert} applies to a wider range of $ \sigma_{\phi h} $, almost up to the maximal value    (\ref{sig_cond}).

Many of these features are also shared by models with the $\phi^4$ inflaton potential. The efficient Higgs production is hindered by the Higgs self--interaction,
however the situation is complicated by the fact that significant trilinear couplings generally induce a non--zero VEV for the dark matter field, which makes it unstable (see \cite{Abolhasani:2009nb} for the simplest example).
The potential becomes a non--trivial function of the 3 fields $\phi, h , s$, whose vacuum structure requires a separate discussion.
 A  careful parameter space analysis is beyond the scope of the present work and will be performed elsewhere.

\section{Challenges for reheating via $\phi^2 h^2 $ interaction}
\label{challenge}

\begin{figure}[t!]
  \includegraphics[width=0.4999\textwidth]{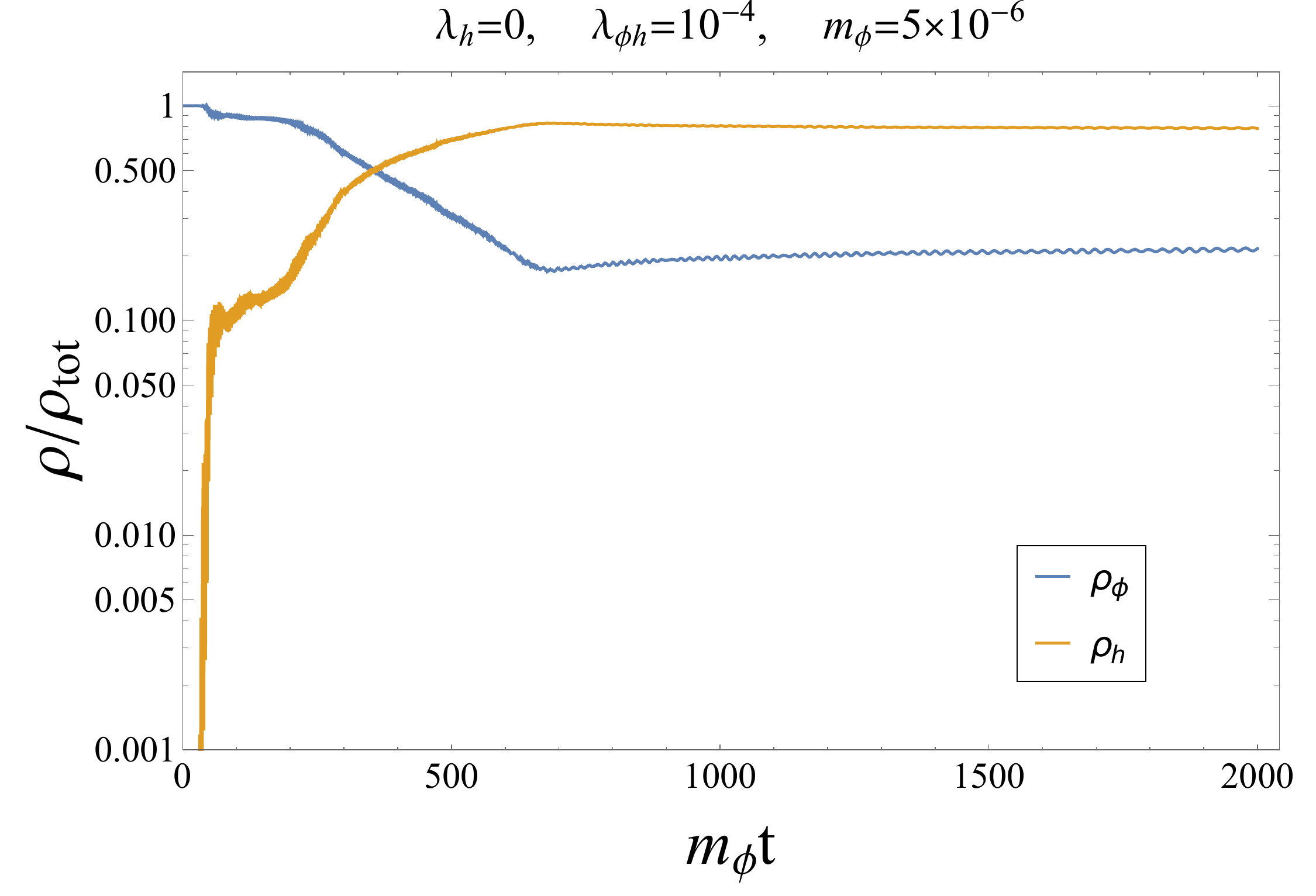}
  \includegraphics[width=0.4999\textwidth]{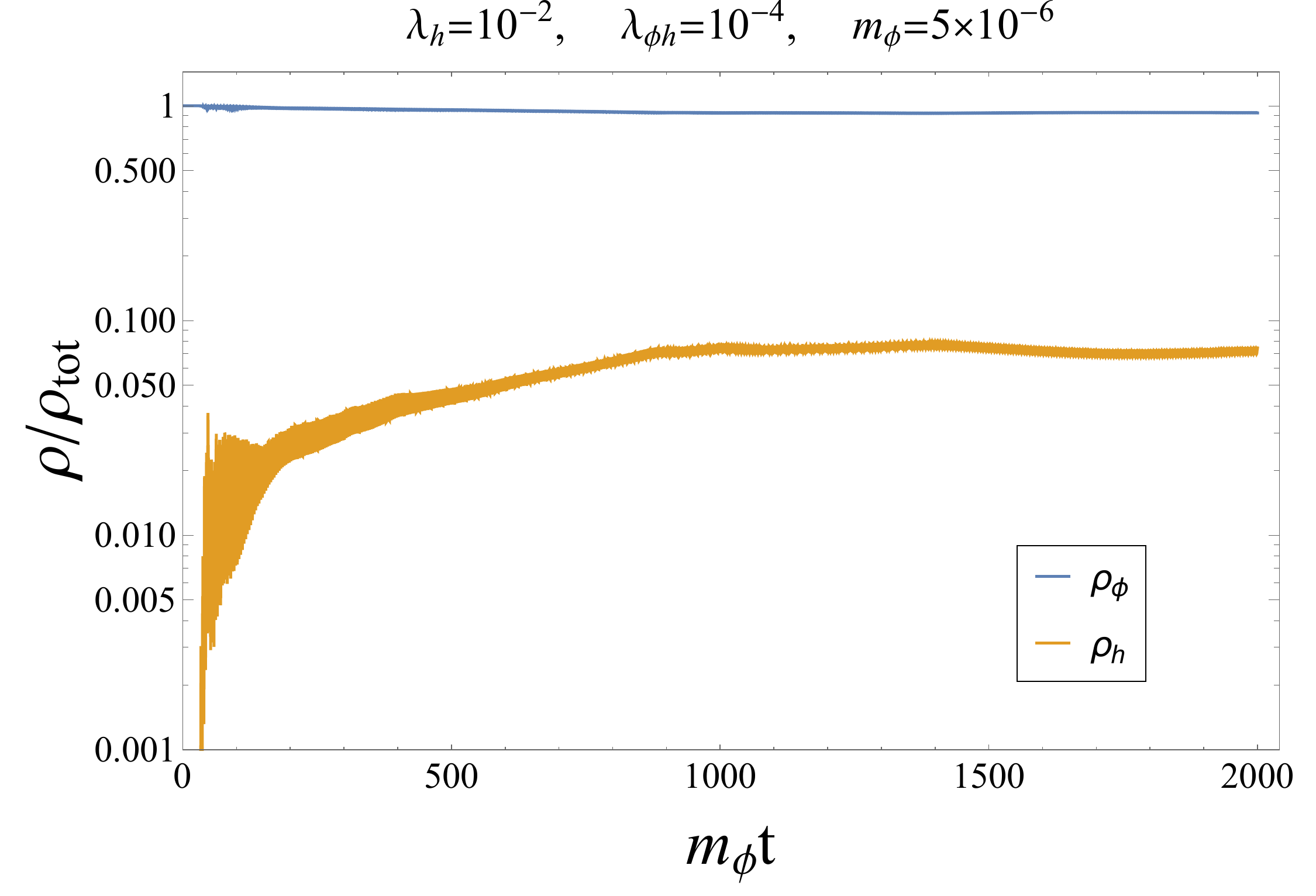}
  \caption{ Energy density of the inflaton and the Higgs (4 d.o.f.) normalized to the total energy density for
  the $\phi^2$ preheating potential with
  $\lambda_{\phi h} =10^{-4}$, $m_\phi = 5\times 10^{-6}$ in Planck units.
  {\it Left:}  no Higgs self--interaction. {\it Right:} $\lambda_h = 10^{-2}$.     Produced with LATTICEEASY.} 
    \label{p2h2}
  \end{figure}

Let us now consider case (c),
 \begin{equation}
V_{\phi h} = {1\over 4} \lambda_{\phi h} \, \phi^2 h^2 ~~,~~ V_{\phi s} =   {1\over 2} \sigma_{\phi s} \, \phi s^2 \;.
\end{equation}
The SM radiation is then produced only during the inflaton oscillation epoch, while DM is generated subsequently via perturbative decay $\phi \rightarrow ss$.
Clearly,  $\lambda_{\phi h}$  has to be sufficiently large  to generate  enough SM matter. Depending on the model, this large coupling may still be consistent with flatness of the inflaton potential.
  In the best case scenario, the efficient Higgs production results in quasi--equilibrium   in the Higgs--inflaton  system (Fig.\,\ref{p2h2}, left panel). This could also extend to the full SM or its subset.
    The energy is then distributed almost democratically among the relativistic degrees of freedom which reach quasi--equilibrium \cite{Lebedev:2021zdh},
  \begin{equation}
  { \rho_\phi \over \rho_{\rm tot} } \sim {1\over \# \;{\rm d.o.f.}}
  \end{equation}
Given that the SM has about 107 d.o.f. at high temperature,
this fraction is bounded {\it from below} by about 1/100  and remains constant in the relativistic regime until inflaton decay becomes important. 
The total number of the inflaton quanta is conserved at this stage and, since the average energy per quantum is roughly the same for all species, a similar relation applies to $n_\phi / n_{\rm tot}$. 

The SM sector thermalizes quickly so that one can use relativistic thermodynamic relations, e.g. $s_{\rm SM} \sim 4 n_{\rm SM}$. The total number of the DM quanta is twice the number of the inflaton quanta,
hence
 \begin{equation}
 Y = {n \over s_{\rm SM}} \sim {n_\phi \over n_{\rm SM}} > {\rm few } \times 10^{-3} \;.
  \end{equation}
 This number far exceeds  the observed value $Y_{\rm obs} = 4 \times 10^{-10} \; {\rm GeV}/m_s $, given that $m_s \gtrsim
10$ keV as required by  the structure formation constraints.\footnote{In our case, dark matter is non--thermal such that the exact bound depends on its momentum distribution. Note that the characteristic energy per quantum in the dark sector is similar to (or greater than) that
in the SM sector and very light DM remains relativistic at the epoch of  structure formation.} 
Therefore, the resulting Universe is dark and unrealistic.

This {\it conservative} lower bound is obtained under the assumption that the energy transfer from the inflaton to the Higgs is very efficient. In reality, the Higgs field has self--interaction 
which leads to  an
effective Higgs mass--squared of order $\lambda_h \langle h^2 \rangle$. As the Higgs variance grows, the Higgs becomes heavy and the resonance terminates. As a result, only a tiny fraction 
of the inflaton energy gets transferred to the Higgses (Fig.\,\ref{p2h2}, right panel), which makes the lower bound on $Y$ much stronger.
In either case, reheating via $\phi^2 h^2 $ for a non--thermal inflaton  appears unrealistic. Similar arguments apply to the $\phi^4$ inflaton potential \cite{Lebedev:2021zdh}.

The above conclusion is however  evaded if the inflaton--Higgs coupling is so large such that the inflaton thermalizes with the Higgs and eventually undergoes freeze--out in the non--relativistic regime.
This scenario will be presented in detail elsewhere \cite{next}.

\section{Conclusion}

We have studied production of scalar dark matter and the Higgs bosons due to their direct couplings to the inflaton. 
The focus of this work is on the effect of collective phenomena such as resonances, backreaction and rescattering
of the produced particles. These make a crucial impact on the non--thermal dark matter abundance.
In particular, we find that the system reaches the state of quasi--equilibrium for the inflaton--DM coupling above a certain value,
which is far below that required for thermalization.
In this case, the dark matter abundance becomes  independent of the couplings and obeys a universal relation (cf.\,Eqs.\,\ref{sigma_simp},\ref{Ystrong}),
which applies to  both $\phi^2 $ and $\phi^4$ preheating inflaton potentials. 
 
 The renormalizable inflaton couplings to dark matter and the Higgs field  are sufficient to {\it fully} describe the reheating and DM production,
 leading to a realistic picture of the Early Universe. The non--perturbative DM production mechanisms are very  efficient such that
 even tiny couplings can generate the correct relic abundance. We delineate the corresponding parameter space for the $\phi^2$ and $\phi^4 $
 inflaton potentials during the preheating epoch.

In this work, we have focussed on the effects of the direct inflaton--DM couplings and assumed that the other sources of dark matter such as the non--minimal scalar coupling to gravity \cite{Fairbairn:2018bsw} and the 
scalar condensate \cite{Enqvist:2014zqa,Markkanen:2018gcw}  are subleading. If these additional sources are significant, our results can be interpreted as the upper bounds on the relevant inflaton couplings.
 \\ \  \\
{\bf Acknowledgements.} The authors wish to thank the Finnish Grid and Cloud Infrastructure (FGCI) for supporting this project with computational and data storage resources.
F.S. and T.S. acknowledge support from the EDUFI program.

 \appendix

\section{Simulation details}

In this work, we have performed{} lattice simulations with CLUSTEREASY, the parallel{} computing version of LATTICEEASY{}. 
For most purposes, the dimension{} of the lattice was set to $D=3$ and the number{} of the grid points per edge was{} fixed at $N=128$ ($128^3$ in total).
The simulations mainly{} target the late time behavior of the system, for which{} the UV momentum spectrum is essential.  To capture the relevant features, 
we have made the upper bound{} of the momentum space $k_{max}$ (in LATTICEEASY convention) dynamical, i.e. coupling--dependent.
For example, in the case of the $\phi^4$ inflaton potential with the $\lambda_{\phi h}$ coupling, the size of the box $L$ in rescaled distance units was set to  
\begin{equation}
L=\frac{\pi \sqrt{D}N}{40} \left( \frac{ \lambda_{\phi h}}{\lambda_{\phi} (4 \pi^2)} \right)^{-0.25} \;,
\end{equation}
such that
\begin{equation}
k_{max}=k_{min} \times  \frac{\sqrt{D}}{2}N = \frac{2\pi}{L}  \frac{\sqrt{D}}{2}N=40 \times  \left( \frac{ \lambda_{\phi h}}{\lambda_{\phi} (4 \pi^2)} \right)^{0.25} \;,
\end{equation}
where $k_{min}$ represents the lower bound of the momentum space in rescaled units and the pre-factor 40 has been  determined empirically.
To verify  reliability of our results{}, we have run extended 2D simulations with{}  $N=1024$ which also capture the relevant infra--red physics. We find that{} 
the late time distributions are indeed{} 
consistent.


\begin{thebibliography}{99}

\bibitem{Linde:1990flp}
A.~D.~Linde,
\textit{Particle physics and inflationary cosmology,}
Contemp. Concepts Phys. \textbf{5} (1990), 1-362,
[\texttt{hep-th/0503203}]

\bibitem{Lebedev:2021xey}
O.~Lebedev,
\textit{The Higgs Portal to Cosmology,}
doi:10.1016/j.ppnp.2021.103881,
[\texttt{2104.03342}].

\bibitem{Kofman:1994rk}
L.~Kofman, A.~D.~Linde and A.~A.~Starobinsky,
\textit{Reheating after inflation,}
Phys. Rev. Lett. \textbf{73} (1994), 3195-3198,
[\texttt{hep-th/9405187}].

\bibitem{Kofman:1997yn}
L.~Kofman, A.~D.~Linde and A.~A.~Starobinsky,
\textit{Towards the theory of reheating after inflation,}
Phys. Rev. D \textbf{56} (1997), 3258-3295,
[\texttt{hep-ph/9704452}].

\bibitem{Greene:1997fu}
P.~B.~Greene, L.~Kofman, A.~D.~Linde and A.~A.~Starobinsky,
\textit{Structure of resonance in preheating after inflation,}
Phys. Rev. D \textbf{56}, 6175-6192 (1997),
[\texttt{hep-ph/9705347}]

\bibitem{Felder:2000hj}
G.~N.~Felder, J.~Garcia-Bellido, P.~B.~Greene, L.~Kofman, A.~D.~Linde and I.~Tkachev,
\textit{Dynamics of symmetry breaking and tachyonic preheating,}
Phys. Rev. Lett. \textbf{87} (2001), 011601,
[\texttt{hep-ph/0012142}]

\bibitem{Dufaux:2006ee}
J.~F.~Dufaux, G.~N.~Felder, L.~Kofman, M.~Peloso and D.~Podolsky,
\textit{Preheating with trilinear interactions: Tachyonic resonance,}
JCAP \textbf{07} (2006), 006,
[\texttt{hep-ph/0602144}]

\bibitem{Khlebnikov:1996mc}
S.~Y.~Khlebnikov and I.~I.~Tkachev,
\textit{Classical decay of inflaton,}
Phys. Rev. Lett. \textbf{77}, 219-222 (1996),
[\texttt{hep-ph/9603378}]

\bibitem{Prokopec:1996rr}
T.~Prokopec and T.~G.~Roos,
\textit{Lattice study of classical inflaton decay,}
Phys. Rev. D \textbf{55}, 3768-3775 (1997),
[\texttt{hep-ph/9610400}]

\bibitem{Felder:2000hq}
G.~N.~Felder and I.~Tkachev,
\textit{LATTICEEASY: A Program for lattice simulations of scalar fields in an expanding universe,}
Comput. Phys. Commun. \textbf{178}, 929-932 (2008),
[\texttt{hep-ph/0011159}]

\bibitem{Heikinheimo:2016yds}
M.~Heikinheimo, T.~Tenkanen, K.~Tuominen and V.~Vaskonen,
\textit{Observational Constraints on Decoupled Hidden Sectors,}
Phys. Rev. D \textbf{94} (2016) no.6, 063506,
[\texttt{1604.02401}]

\bibitem{Heurtier:2017nwl}
L.~Heurtier,
\textit{The Inflaton Portal to Dark Matter,}
JHEP \textbf{12} (2017), 072,
[\texttt{1707.08999}]

\bibitem{McDonald:2001vt}
J.~McDonald,
\textit{Thermally generated gauge singlet scalars as selfinteracting dark matter,}
Phys. Rev. Lett. \textbf{88} (2002), 091304,
[\texttt{hep-ph/0106249}]

\bibitem{Hall:2009bx}
L.~J.~Hall, K.~Jedamzik, J.~March-Russell and S.~M.~West,
\textit{Freeze-In Production of FIMP Dark Matter,}
JHEP \textbf{03} (2010), 080,
[\texttt{0911.1120}]

\bibitem{Lebedev:2019ton}
O.~Lebedev and T.~Toma,
\textit{Relativistic Freeze-in,}
Phys. Lett. B \textbf{798} (2019), 134961,
[\texttt{1908.05491}]

\bibitem{Buttazzo:2013uya}
D.~Buttazzo, G.~Degrassi, P.~P.~Giardino, G.~F.~Giudice, F.~Sala, A.~Salvio and A.~Strumia,
\textit{Investigating the near-criticality of the Higgs boson,}
JHEP \textbf{12} (2013), 089,
[\texttt{1307.3536}]

\bibitem{Lebedev:2012sy}
O.~Lebedev and A.~Westphal,
\textit{Metastable Electroweak Vacuum: Implications for Inflation,}
Phys. Lett. B \textbf{719} (2013), 415-418,
[\texttt{1210.6987}]

\bibitem{Ema:2017ckf}
Y.~Ema, M.~Karciauskas, O.~Lebedev, S.~Rusak and M.~Zatta,
\textit{Higgs\textendash{}inflaton mixing and vacuum stability,}
Phys. Lett. B \textbf{789} (2019), 373-377,
[\texttt{1711.10554}]

\bibitem{Kost:2021rbi}
J.~Kost, C.~S.~Shin and T.~Terada,
\textit{Massless Preheating and Electroweak Vacuum Metastability,}
[\texttt{2105.06939}]

\bibitem{Bezrukov:2007ep}
F.~L.~Bezrukov and M.~Shaposhnikov,
\textit{The Standard Model Higgs boson as the inflaton,}
Phys. Lett. B \textbf{659} (2008), 703-706,
[\texttt{0710.3755}]

\bibitem{Garcia-Bellido:2008ycs}
J.~Garcia-Bellido, D.~G.~Figueroa and J.~Rubio,
\textit{Preheating in the Standard Model with the Higgs-Inflaton coupled to gravity,}
Phys. Rev. D \textbf{79} (2009), 063531,
[\texttt{0812.4624}]

\bibitem{Linde:1983gd}
A.~D.~Linde,
\textit{Chaotic Inflation,}
Phys. Lett. B \textbf{129} (1983), 177-181.

\bibitem{Planck:2018jri}
Y.~Akrami \textit{et al.} [Planck],
\textit{Planck 2018 results. X. Constraints on inflation,}
Astron. Astrophys. \textbf{641} (2020), A10,
[\texttt{1807.06211}]

\bibitem{Figueroa:2016wxr}
D.~G.~Figueroa and F.~Torrenti,
\textit{Parametric resonance in the early Universe\textemdash{}a fitting analysis,}
JCAP \textbf{02} (2017), 001,
[\texttt{1609.05197}]

\bibitem{Podolsky:2005bw}
D.~I.~Podolsky, G.~N.~Felder, L.~Kofman and M.~Peloso,
\textit{Equation of state and beginning of thermalization after preheating,}
Phys. Rev. D \textbf{73} (2006), 023501,
[\texttt{hep-ph/0507096}]

\bibitem{Antusch:2020iyq}
S.~Antusch, D.~G.~Figueroa, K.~Marschall and F.~Torrenti,
\textit{Energy distribution and equation of state of the early Universe: matching the end of inflation and the onset of radiation domination,}
Phys. Lett. B \textbf{811} (2020), 135888,
[\texttt{2005.07563}]

\bibitem{Planck:2015fie}
P.~A.~R.~Ade \textit{et al.} [Planck],
\textit{Planck 2015 results. XIII. Cosmological parameters,}
Astron. Astrophys. \textbf{594} (2016), A13,
[\texttt{1502.01589}]

\bibitem{Arcadi:2019oxh}
G.~Arcadi, O.~Lebedev, S.~Pokorski and T.~Toma,
\textit{Real Scalar Dark Matter: Relativistic Treatment,}
JHEP \textbf{08} (2019), 050,
[\texttt{1906.07659}]

\bibitem{Figueroa:2020rrl}
D.~G.~Figueroa, A.~Florio, F.~Torrenti and W.~Valkenburg,
\textit{The art of simulating the early Universe -- Part I,}
JCAP \textbf{04} (2021), 035,
[\texttt{2006.15122}]

\bibitem{Figueroa:2021yhd}
D.~G.~Figueroa, A.~Florio, F.~Torrenti and W.~Valkenburg,
\textit{CosmoLattice,}
[\texttt{2102.01031}]
   
\bibitem{Dolgov:1989us}
A.~D.~Dolgov and D.~P.~Kirilova,
\textit{ON PARTICLE CREATION BY A TIME DEPENDENT SCALAR FIELD,}
Sov. J. Nucl. Phys. \textbf{51}, 172-177 (1990).
 
\bibitem{Traschen:1990sw}
J.~H.~Traschen and R.~H.~Brandenberger,
\textit{Particle Production During Out-of-equilibrium Phase Transitions,}
Phys. Rev. D \textbf{42}, 2491-2504 (1990).
 
\bibitem{Ichikawa:2008ne}
K.~Ichikawa, T.~Suyama, T.~Takahashi and M.~Yamaguchi,
\textit{Primordial Curvature Fluctuation and Its Non-Gaussianity in Models with Modulated Reheating,}
Phys. Rev. D \textbf{78}, 063545 (2008),
[\texttt{0807.3988}]

\bibitem{Turner:1983he}
M.~S.~Turner,
\textit{Coherent Scalar Field Oscillations in an Expanding Universe,}
Phys. Rev. D \textbf{28} (1983), 1243. 

\bibitem{Hannestad:2004px}
S.~Hannestad,
\textit{What is the lowest possible reheating temperature?,}
Phys. Rev. D \textbf{70} (2004), 043506,
[\texttt{astro-ph/0403291}]
    
\bibitem{DeRomeri:2020wng}
V.~De Romeri, D.~Karamitros, O.~Lebedev and T.~Toma,
\textit{Neutrino dark matter and the Higgs portal: improved freeze-in analysis,}
JHEP \textbf{10} (2020), 137,
[\texttt{2003.12606}]
    
\bibitem{Abolhasani:2009nb}
A.~A.~Abolhasani, H.~Firouzjahi and M.~M.~Sheikh-Jabbari,
\textit{Tachyonic Resonance Preheating in Expanding Universe,}
Phys. Rev. D \textbf{81} (2010), 043524,
[\texttt{0912.1021}]
    
\bibitem{Lebedev:2021zdh}
O.~Lebedev and J.~H.~Yoon,
\textit{Challenges for Inflaton Dark Matter,}
[\texttt{2105.05860}]

\bibitem{next}
O.~Lebedev, T.~Nerdi, T.~Solomko and J.~Yoon,
\textit{Inflaton freeze-out},
to appear.    

\bibitem{Fairbairn:2018bsw}
M.~Fairbairn, K.~Kainulainen, T.~Markkanen and S.~Nurmi,
\textit{Despicable Dark Relics: generated by gravity with unconstrained masses,}
JCAP \textbf{04} (2019), 005,
[\texttt{1808.08236}]
    
\bibitem{Enqvist:2014zqa}
K.~Enqvist, S.~Nurmi, T.~Tenkanen and K.~Tuominen,
\textit{Standard Model with a real singlet scalar and inflation,}
JCAP \textbf{08} (2014), 035,
[\texttt{1407.0659}]

\bibitem{Markkanen:2018gcw}
T.~Markkanen, A.~Rajantie and T.~Tenkanen,
\textit{Spectator Dark Matter,}
Phys. Rev. D \textbf{98} (2018) no.12, 123532,
[\texttt{1811.02586}]

\end{thebibliography}
\end{document}